\documentclass[english,prd,nofootinbib,preprintnumbers,twocolumn,showpacs]{revtex4}
\usepackage[latin1]{inputenc}
\usepackage{amsmath}
\usepackage{amsfonts}
\usepackage{appendix}
\usepackage{amssymb}
\usepackage{epsfig}
\usepackage{graphics,psfrag,rotating}
\usepackage{graphicx}
\usepackage{dcolumn}
\usepackage{bm}
\usepackage{epstopdf}
\usepackage{color}
\usepackage[usenames,dvipsnames,svgnames]{xcolor}
\usepackage[colorlinks=true,linkcolor=red,urlcolor=gray,citecolor=blue]{hyperref}
\usepackage{hyperref}
\usepackage[T1]{fontenc}
\usepackage{multirow}
\usepackage{float}
\usepackage{hyperref}
\usepackage{subfigure}
\newcommand{\Od}{{\cal O}}
\usepackage{enumitem}

\def\3nab{\tilde{\nabla}}

\def\be {\begin{equation}}
\def\ee {\end{equation}}
\def\ba {\begin{align}}
\def\ea {\end{align}}

\def\bc {\begin{center}}
\def\ec {\end{center}}
\def\case#1/#2{\frac{#1}{#2}}

\newcommand{\bver}{\begin{verbatim}}
\newcommand{\ever}{\end{verbatim}}

\newcommand{\bea}{\begin{eqnarray}}
\newcommand{\eea}{\end{eqnarray}}
\newcommand{\beaa}{\begin{eqnarray*}}
\newcommand{\eeaa}{\end{eqnarray*}}

\def\case#1/#2{\textstyle\frac{#1}{#2}}

\begin{document}
\title{Parametrizing modified gravities with vector degrees of freedom: \\
anisotropic growth and lensing}

\author{
Miguel Aparicio Resco 
}
\email{migueapa@ucm.es}
\affiliation{Departamento de F\'{\i}sica Te\'orica and UPARCOS, Universidad Complutense de Madrid, 28040 
Madrid, Spain}
\author{Antonio L.\ Maroto}
\email{maroto@ucm.es}
\affiliation{Departamento de F\'{\i}sica Te\'orica and UPARCOS, Universidad Complutense de Madrid, 28040 
Madrid, Spain}




\begin{abstract} 

We consider the problem of parametrizing modified gravity theories that include  an
additional vector field in the sub-Hubble regime within the quasi-static approximation. We start from the most general set of second order equations
for metric and vector field perturbations and allow for both temporal and spatial components of the background 
vector field. We find that in the case in which dark matter obeys standard 
conservation equations, eight parameters are needed to fully characterize
the theory. If dark matter vorticity can be neglected, the number of independent parameters is reduced to four. In addition to the 
usual scale and redshift dependence, the effective parameters 
have an additional angular dependence induced by 
the preferred direction set by the background vector. In the considered
sub-Hubble regime, we show that 
this angular dependence appears only through even multipoles and generates anisotropies
in the growth function which translate into anisotropies in the galaxy and 
lensing convergence power spectra. The angular dependence 
generated by the preferred direction is different from that induced 
by redshift space distortions and could be disentangled in the data collected
by future galaxy surveys.

\end{abstract} 


\maketitle


\section{Introduction}

In the next decade a new suite of cosmological surveys such as J-PAS \cite{JPAS}, DESI \cite{DESI}, Euclid \cite{Euclid}, TAIPAN \cite{TAIPAN}, LSST \cite{LSST}, WALLABY or SKA \cite{SKA} will provide us with unprecedented amount of data on  the distribution, shapes and peculiar velocities of galaxies on
very large scales. 
These observations could shed new light not only on the nature of dark energy, responsible 
for the accelerated expansion of the universe, but also on the behaviour of 
the gravitational interaction on cosmological scales. Different forecast analysis
suggest that sensitivities in the percentage level for the equation of 
state of dark energy and for certain parameters of modified gravity theories 
at different redshifts will be attainable by the mentioned surveys \cite{SKA,forecast}. 

On the theory side,  models for dark energy or modified gravity have been proposed in the last years aiming at exploring consistent alternatives to the standard $\Lambda$CDM cosmology \cite{DE, Clifton}. Given the plethora of viable
cosmologies we have to date, it is becoming more and more indispensable to have tools that allow to confront alternative cosmologies with data
in a model independent way.
This fact has triggered the development of effective theoretical descriptions which 
have the ability to capture general  modifications of gravity in a few parameters which can be tested by observations \cite{Clifton,Pogosian}. Thus, in the relevant range of scales for observations of structure growth and weak lensing, i.e. in the sub-Hubble regime  where the so called quasi-static approximation (QSA) can be safely employed, 
and provided dark matter satisfies standard conservation equations, all the relevant information of a very general class of theories of gravity involving additional scalar degrees of freedom can be encapsulated in 
only two parameters, namely an effective Newton constant $\mu(a,k)=G_{\text{eff}}/G$ and a gravitational slip parameter $\gamma(a,k)$ \cite{Pogosian,Clifton}. 
Any deviation with respect to $\mu=\gamma=1$ could signal a breakdown of
General Relativity (GR).

In the scalar case, the so called beyond-Horndeski theories \cite{Hordenski} provide the most general expression for a  local and covariant scalar theory in four dimensions with no higher than second order derivatives in their equations of motion. Explicit expression for the effective parameters have been derived for such theories in \cite{HordParam0,HordParam1,HordParam2}.

Modified gravities with extra vector degrees of freedom have also been 
widely studied in recent years.  Since the first proposal in \cite{Ford}, different models of vector inflation
have been  studied in the literature \cite{Golovnev,Koivisto,Joda1,Joda2}.
Vector models for dark energy based on massive \cite{ArmendarizPicon,Boehmer} or 
massless \cite{Beltran1,Beltran2,Beltran3} vector fields have also been proposed.
Vector dark matter based on hidden sector gauge bosons have been analyzed in \cite{Redondo}, and the role of vectors in the generation of metric perturbations in the so called curvaton scenario
has also been considered in \cite{Dimopoulos}.  Oscillating massive vector
fields were considered as  non-thermal dark matter candidates in \cite{Nelson} and ultra-light vector dark matter models have been explored in \cite{ULV}. 
The most general framework considered so far for modifications of 
gravity induced by vectors is given by the so called beyond generalized Proca 
models \cite{BGP,BGP2} which propagate three degrees of freedom (two transverse and 
one longitudinal) corresponding to a massive spin 1 field.

Most of the work done so far on modified theories with vector degrees of freedom  concentrate in the case in which the background vector field is purely temporal. In that case, it can
 be shown \cite{DeFelice:2016uil, Lagos} that very much as in 
 the scalar case,  the theory can be parametrized by the same two phenomenological 
 parameters $\mu$ and $\gamma$  in the quasi-static approximation. However, the possibility of having spatial 
 components in the background cannot be excluded a priori, provided their effect
 on the expansion is small \cite{isotropyPlanck}. As a matter of fact
 the existence of a preferred spatial direction in the universe has been advocated as a possible explanation of the low-multipole anomalies of the CMB \cite{anomalies,Campanelli,MDE}.

The aim of this work is to extend previous results in two directions, on 
one hand we will work at the level of equations of motions and consider the most general modified second order equations for perturbations involving vector fields, and on the other, we will consider a general background vector field 
involving both temporal and spatial components. The presence of the 
background spatial components implies that the number of effective parameters needed to parametrize the theory in the QSA is increased to eight. This number can be reduced to four
in the case in which dark matter vorticity could be neglected. Moreover,  the spatial background vector
generates an additional angular dependence
in the effective parameters. In particular,  $\mu= \mu(a,k,x)$ and $\gamma=\gamma(a,k,x)$ with $x=\hat k \cdot \hat A$ given by the angle between the 
wavevector $\vec k$ and 
the preferred background direction $\hat A$. This fact opens up the 
possibility of generating anisotropic growth of structures and anisotropies
in the weak lensing signal. We will consider linear scales and assume that 
possible screening mechanisms only work at much smaller distances.  

The potential observable effects of a preferred direction in  cosmology  have been previously analyzed in the literature, both 
on the CMB temperature power spectrum 
\cite{ACW, isotropyPlanck} and  in the matter distribution 
 in \cite{Pullen,Jeong,Shiraishi}. Also very recently the possible impact of  vector (vorticity) perturbations with anisotropic power spectrum on the galaxy distribution has been analyzed in \cite{Durrer2}.
 
  In those works, the anisotropy is assumed to be present only in the  primordial  power spectra, which can be  generated for instance in models of inflation with vectors \cite{Bartolo,Bartolo2,Joda3} or higher-spin 
fields \cite{Bartolo3}, but not in the 
transfer functions which are assumed 
isotropic. However, in modified gravities with a spatial vector background, anisotropies could be 
induced in the transfer functions themselves, thus affecting both 
galaxy and lensing convergence power spectra.  Such anisotropies have a different angular dependence from those generated by redshift space distortions
and could be disentangled in the data of future galaxy surveys. 

The paper is organized as follows: in section \ref{sec1} we summarize the results for the scalar field case in \cite{Pogosian}, and in section \ref{sec2} we analyze a particular case of a non-minimally coupled scalar field; in section \ref{sec3} we consider the general case for the linearized equations of perturbations with a vector field in the background, we define the modified gravity parameters and we find general expressions for them. Then, in \ref{sec4}, we consider a particular case of a non-minimally coupled vector field and we obtain simple equations for this modified gravity parameters in the sub-Hubble regime which follow the general framework. 
In \ref{sec5} we explore the possible observational consequences of these
theories in the galaxy and lensing power spectra. Finally in section \ref{sec6} we briefly discuss the results and conclusions.

\section{General scalar field case}\label{sec1}
In order to introduce the formalism we will use in the rest of the paper, 
we will review the well-known scalar case following the approach in \cite{Pogosian}. 

Let us consider a modification of General Relativity 
which involves an additional scalar field $\phi$. We will study the scalar 
perturbed flat Robertson-Walker metric in the longitudinal gauge,
\begin{eqnarray}\label{s1.1}
ds^2=a^2(\tau) \, [-(1+2\Psi)d\tau^2+(1-2\Phi)d{\bf x}^2],
\end{eqnarray}
and the scalar field is also perturbed as $\phi=\phi_0(\tau)+\delta \phi$, 
The modified Einstein equations at the perturbation level are,
\begin{equation}\label{s1.2}
\delta \bar{G}^{\mu}_{\,\,\,\nu}=8 \pi G \, \delta T^{\mu}_{\,\,\,\nu},
\end{equation}
where the perturbed modified Einstein tensor $\delta \bar{G}^{\mu}_{\,\,\,\nu}$  can depend on both the metric potentials
$\Phi$, $\Psi$ and the perturbed field $\delta\phi$ to first order. The only matter-energy content relevant at late times is pressureless matter so that,
\begin{equation}\label{s1.3}
\delta T^{0}_{\,\,\,0}=-\rho \, \delta,
\end{equation}
\begin{equation}\label{s1.4}
\delta T^{0}_{\,\,\,i}=-\rho \, v_{i},
\end{equation}
\begin{equation}\label{s1.5}
\delta T^{i}_{\,\,\,j}=0,
\end{equation}
where $v_{i}$ is the three-velocity of matter, $\rho$ is the density and $\delta$ the density contrast.

In components, we have, a priori, the following modified Einstein equations corresponding to  $\delta \bar{G}^{0}_{\,\,0}$, $\delta \bar{G}^{0}_{\,\,i}$ and $\delta \bar{G}^{j}_{\,\,i}$, but not all of them are independent because of the  Bianchi identities. Thus, imposing $\nabla_{\mu}\bar{G}^{\mu}_{\,\,\nu}=0$ at the perturbation level and in the sub-Hubble regime, we find $\partial_{i} \delta \bar{G}^{i}_{\,\,\nu}=0$, so that in the Fourier space we have,
\begin{align}\label{s1.6}
\hat{k}^{i} \delta \bar{G}^{0}_{\,\,i}=0, \,\,\,\,\, \hat{k}^{i} \delta \bar{G}^{j}_{\,\,i}=0,
\end{align}
being $\hat{k}_{i} = k_i / k$. Taking these restrictions into account we have only two independent equations which we take as those corresponding to $\delta \bar{G}^{0}_{\,\,0}$ and $\delta \bar{G}^{i}_{\,\,i}$. Additionally, we have the equation of motion for the scalar field $\phi$ that provides a third equation for the perturbation $\delta \phi$, so that we can write in general the following system of equations for first order perturbations
\begin{eqnarray}\label{s1.7}
a_{1 1} \, \Psi + a_{1 2} \, \Phi + a_{1 3} \, \delta \phi = -4 \pi G a^2\, \rho \, \delta,
\end{eqnarray}
\begin{eqnarray}\label{s1.8}
a_{2 1} \, \Psi + a_{2 2} \, \Phi + a_{2 3} \, \delta \phi = 0,
\end{eqnarray}
\begin{eqnarray}\label{s1.9}
a_{3 1} \, \Psi + a_{3 2} \, \Phi + a_{3 3} \, \delta \phi = 0,
\end{eqnarray}
where $a_{i j}$ are general differential operators although we will restrict ourselves  to  second order operators and we have written $4\pi G$ on the r.h.s. for convenience to compare with the standard Poisson equation. Then, we introduce the quasi-static approximation  in which we neglect all time derivatives of perturbations so 
that, in Fourier space, equations (\ref{s1.7}) - (\ref{s1.9}) are just algebraic equations for $(\Phi, \,\Psi, \,\delta \phi)$ in terms of $\delta$. 
Notice that the first Bianchi condition in (\ref{s1.6}) implies that the 
velocity perturbation $v$ does not contribute to the equations of motion in the 
sub-Hubble regime. We can now eliminate the 
scalar degree of freedom simply solving for $\delta \phi$ in (\ref{s1.9}) and 
substituting in (\ref{s1.7}) and (\ref{s1.8}), obtaining in this way
the effective equations for the metric perturbations, which in general can be written as
\begin{eqnarray}\label{s1.12}
k^2 \, \Phi = -4\pi G\,a^2 \, \mu \, \gamma \, \rho\, \delta,
\end{eqnarray}
\begin{eqnarray}\label{s1.13}
k^2 \, \Psi = -4\pi G\,a^2 \, \mu\, \rho \,\delta.
\end{eqnarray}
Notice that in the sub-Hubble regime
$\delta$ agrees with the density perturbation $\Delta$ used in \cite{Pogosian} since $\Delta=\delta+\frac{3 {\cal H} v}{k}$. 

Thus we see that in the QSA, the modified Einstein equations can be parametrized in terms of 
two functions of time and scale $\mu(a,k)$ and $\gamma(a,k)$. These parameters 
can be understood  as an effective Newton constant $G_{\text{eff}}(a,k)$ given by
\begin{eqnarray}\label{s1.16}
\mu(a,k)=\frac{G_{\text{eff}}}{G},
\end{eqnarray}
which modifies the standard Poisson equation, and the so called gravitational slip parameter  
\begin{eqnarray}\label{s1.17}
\gamma(a,k)=\frac{\Phi}{\Psi}, 
\end{eqnarray}
which in general modifies the equation for the lensing potential $(\Phi+\Psi)/2$.

In order to obtain the explicit $k$ dependence of $\mu$ and $\gamma$ 
we go back to equations (\ref{s1.7}) - (\ref{s1.9}). As mentioned 
before in the QSA each $a_{i j}$ 
coefficient is a general second order operator of the spatial derivatives, 
 so that in  Fourier space
they can be written as,
\begin{eqnarray}\label{s1.18}
a_{i j}(a,k)= d_{i j}(a) + c_{i j}(a) \, k^2,
\end{eqnarray}
where $d_{i j}(a)$ and $c_{i j}(a)$ are general functions of  background quantities. A priori, this implies  on dimensional grounds  that $d_{ij}$ will be of order ${\cal H}^2$ with ${\cal H}=a'/a$ the Hubble parameter (where prime denotes derivative
with respect to $\tau$) and therefore can be neglected compared to the the $c_{ij}$ terms in 
the sub-Hubble regime. Notice 
however that $d_{i j}$ could also involve mass terms for the scalar fields
which could be larger than ${\cal H}^2$ and therefore we will keep 
them for the sake of generality.  Also, we assume that in general the $c_{ij}$ coefficients 
are of order $\Od({\cal H}^0)$. In such a case the sound horizon of perturbations agrees with the Hubble horizon. In the case in which such coefficients are suppressed then our results
will be valid only below the sound horizon, down to the non-linearity scale. Thus we can write
\begin{widetext}
\begin{eqnarray}\label{s1.19.1}
(d_{11} + c_{11} \, k^2) \, \Psi + (d_{12} + c_{12} \, k^2) \, \Phi + (d_{13} + c_{13} \, k^2) \, \delta \phi = -4\pi G\,a^2 \rho \delta,
\end{eqnarray}
\begin{eqnarray}\label{s1.20.1}
(d_{21} + c_{21} \, k^2) \, \Psi +(d_{22} + c_{22} \, k^2) \, \Phi + (d_{23} + c_{23} \, k^2) \, \delta \phi = 0,
\end{eqnarray}
\begin{eqnarray}\label{s1.21.1}
(d_{31} + c_{31} \, k^2) \, \Psi + (d_{32} + c_{32} \, k^2) \, \Phi + (d_{33} + c_{33} \, k^2) \, \delta \phi = 0.
\end{eqnarray}
\end{widetext}
 Then, using equations (\ref{s1.16}) and (\ref{s1.17}) we can solve the previous system and we obtain,
\begin{eqnarray}\label{s1.22.1}
\mu(a,k)=\frac{k^2 \, (1 + p_1(a)k^2 + p_2(a)k^4)}{p_3(a) + p_4(a)k^2 + p_5(a)k^4 + p_6(a)k^6},
\end{eqnarray}
and 
\begin{eqnarray}\label{s1.23.1}
\gamma(a,k)=\frac{p_7(a) + p_8(a)k^2 + p_9(a)k^4}{1 + p_1(a)k^2 + p_2(a)k^4},
\end{eqnarray}
where $p_i(a)$ are  functions of the $c_{ij}$ and $d_{ij}$ coefficients. 
In order to compare with previous results, we will neglect all the mass terms except for $d_{33}$ as is customarily done 
in modified gravity theories \cite{Pogosian,Tsujikawa:2008uc}
\begin{eqnarray}\label{s1.19}
c_{11} \, k^2 \, \Psi + c_{12} \, k^2 \, \Phi + c_{13} \, k^2 \, \delta \phi = -4\pi G\,a^2 \rho \delta,
\end{eqnarray}
\begin{eqnarray}\label{s1.20}
c_{21} \, \Psi + c_{22} \, \Phi + c_{23} \, \delta \phi = 0,
\end{eqnarray}
\begin{eqnarray}\label{s1.21}
c_{31} \, k^2 \, \Psi + c_{32} \, k^2 \, \Phi + (d_{33} + c_{33} \, k^2) \, \delta \phi = 0,
\end{eqnarray}
in this situation we obtain simple expressions for $\mu$ and $\gamma$,
\begin{eqnarray}\label{s1.22}
\mu(a,k)=\frac{1 + p_{1}(a)k^2}{p_{2}(a) + p_{3}(a)k^2},
\end{eqnarray}
and 
\begin{eqnarray}\label{s1.23}
\gamma(a,k)=\frac{p_{4}(a) + p_{5}(a)k^2}{1 + p_{1}(a)k^2}.
\end{eqnarray}
Finally, in the case in which the comoving mass of the scalar field is also of order ${\cal H}$ we can neglect $d_{33}$ and obtain,
\begin{eqnarray}\label{s1.24}
\mu(a,k)=p_{6}(a),
\end{eqnarray}
\begin{eqnarray}\label{s1.25}
\gamma(a,k)=p_{7}(a),
\end{eqnarray}
so that the effective parameters are $k$-independent.

Additionally, if we consider one scalar field more, i.e. $(\phi_1,\phi_2)$ then the effective equations would read
\begin{align}\label{s1.26}
c_{11} \, k^2 \, \Psi +& c_{12} \, k^2 \, \Phi + c_{13} \, k^2 \, \delta \phi_1 \nonumber\\
&+ c_{14} \, k^2 \, \delta \phi_2 = -4\pi G\,a^2 \rho \delta,
\end{align}
\begin{align}\label{s1.27}
c_{21} \, \Psi + c_{22} \, \Phi + c_{23} \, \delta \phi_1 + c_{24} \, \delta \phi_2 = 0,
\end{align}
\begin{align}\label{s1.28}
c_{31} \, k^2 \, \Psi + c_{32} \, k^2 \, \Phi +& (d_{33} + c_{33} \, k^2) \, \delta \phi_1 \nonumber\\
&+ (d_{34} + c_{34} \, k^2) \, \delta \phi_2 = 0,
\end{align}
\begin{align}\label{s1.29}
c_{41} \, k^2 \, \Psi + c_{42} \, k^2 \, \Phi +& (d_{43} + c_{43} \, k^2) \, \delta \phi_1 \nonumber\\
&+ (d_{44} + c_{44} \, k^2) \, \delta \phi_2 = 0,
\end{align}
and we obtain,
\begin{eqnarray}\label{s1.30}
\mu(a,k)=\frac{1 + p_1(a)k^2 + p_2(a)k^4}{p_3(a) + p_4(a)k^2 + p_5(a)k^4},
\end{eqnarray}
\begin{eqnarray}\label{s1.31}
\gamma(a,k)=\frac{p_6(a) + p_7(a)k^2 + p_{8}(a)k^4}{1 + p_1(a)k^2 + p_2(a)k^4}.
\end{eqnarray}
Thus we see that compared to (\ref{s1.22}) and (\ref{s1.23}), each scalar degree of freedom adds an extra even power term in $k$ in the numerator and denominator.  

In the following section we apply these general results to a particular non-minimally coupled scalar field model. 

\section{Non-minimally coupled scalar field}\label{sec2}

Let us consider the following action for a non-minimally coupled scalar field,
\begin{align}\label{s2.1}
S = -\,\int \, d^4x \, \frac{1}{2} \, \sqrt{-g} \, \left[ g^{\mu \nu} \, \partial_{\mu}\phi \, \partial_{\nu}\phi + (m^2 + \xi \, R) \, \phi^2 \right],
\end{align}
where $\xi$ is the dimensionless non-minimal coupling constant, $R$ is the Ricci scalar, and $m$ is the mass of the scalar field. The energy-momentum tensor for this scalar field is,
\begin{align}\label{s2.2}
T_{\mu \nu}^{(s)} =& (1-2 \, \xi) \, \partial_\mu \phi \, \partial_\nu \phi + \left( 2 \xi - \frac{1}{2} \right) \, g_{\mu \nu} \, g^{\alpha \beta} \, \partial_\alpha \phi \, \partial_\beta \phi \nonumber\\
&-2 \, \xi \, \phi \, \nabla_\nu \nabla_\mu \phi - \frac{1}{2} \, (1-3 \xi) \, m^2 \, g_{\mu \nu} \, \phi^2 \nonumber\\
& + \xi \, (G_{\mu \nu} + \frac{3}{2} \, \xi \, R \, g_{\mu \nu}) \, \phi^2 + \frac{1}{2} \, \xi \, g_{\mu \nu} \, \phi \, \Box \phi,
\end{align}
being $\Box \equiv g^{\alpha \beta} \, \nabla_{\alpha} \nabla_{\beta}$ and the total energy-momentum tensor reads,
\begin{align}\label{s2.3}
T_{\mu \nu}^{(t)} = T_{\mu \nu}^{(m)} + T_{\mu \nu}^{(s)},
\end{align}
being $T_{\mu \nu}^{(m)}$ the energy-momentum tensor of pressureless matter.  We consider the metric (\ref{s1.1}) and  perturb the Einstein equations so that
\begin{eqnarray}
\delta G_{\mu\nu}=8\pi G \,\delta T_{\mu\nu}^{(t)}.
\end{eqnarray}
Thus using the procedure described in  previous section we obtain the system (\ref{s1.19} - \ref{s1.21}). As mentioned before, we will neglect all the mass terms except for $d_{33}$, so that

\begin{widetext}

\begin{align}\label{s2.4}
 - \frac{3}{2} \, \xi^2 \phi_p^2 \, k^2 \, \Psi + (1-\xi \phi_p^2 [1-3\xi]) \, k^2 \, \Phi + \frac{1}{4} \, \xi \phi_p^3 \, k^2 \, \delta \phi_p = -4\pi G\,a^2 \rho \delta,
\end{align}
\begin{align}\label{s2.5}
\left( \xi \phi_p^2 \left[ 1- \frac{9}{2} \xi \right] - 1 \right) \, \Psi - \left( \xi \phi_p^2 \left[ 1- 9 \xi \right] - 1 \right) \, \Phi - \frac{1}{4} \, \xi \phi_p^3 \, \delta \phi_p = 0,
\end{align}

\begin{align}\label{s2.6}
2 \, \xi (3 \xi + \bar{\phi} [1-3\xi]) \, k^2 \, \Psi - 4 \, \xi (3 \xi + \bar{\phi} [1-3\xi]) \, k^2 \, \Phi + \left\{ \left(3 \xi + \bar{\phi} \left[1-\frac{3}{2}\xi\right]\right) \, k^2 + \left[1-3\xi\right] \, \bar{\phi} \, a(\tau)^2 \, m^2 \right\} \, \phi_p \, \delta \phi_p = 0,
\end{align}

\end{widetext}

being,

\begin{align}\label{s2.7}
\phi_p \equiv \frac{\phi_0 (\tau)}{M_p},  \,\,\, \delta\phi_p \equiv \frac{\delta \phi}{M_p}, \,\,\, \bar{\phi} \equiv \frac{\phi_0'(\tau)}{H(\tau) \phi_0(\tau)}.
\end{align}
Let us consider the minimal  ($\xi = 0$) and non-minimal coupling ($\xi \neq 0$) cases.
\\
\subsection*{a) Minimal coupling:}

We substitute $\xi=0$ in equations (\ref{s2.4})-(\ref{s2.6}), we solve the system and using (\ref{s1.16}) and (\ref{s1.17}) we obtain $\mu$ and $\gamma$. We find trivially that $\mu=\gamma=1$ in this situation as expected \cite{Tsujikawa:2008uc}.

\subsection*{b) Non-minimal coupling:}

Now we consider $\xi \neq 0$ and we obtain expressions for $\mu$ and $\gamma$ which follow the general form of (\ref{s1.22}) and (\ref{s1.23}) being,
\begin{eqnarray}
p_{i}(a) = \frac{f_i(a)}{f(a)  a^2(\tau) m^2}, \; i=1\dots 5
\end{eqnarray}

with,
\begin{widetext}

\begin{eqnarray}
f(a) = 54 \phi_p^2 \bar{\phi} \xi^3 - 24 \phi_p^2 \bar{\phi} \xi^2 + 2 \bar{\phi} (\phi_p^2 + 3) \xi - 2 \bar{\phi},
\end{eqnarray}

\begin{eqnarray}
f_1(a) = (21 \bar{\phi}-48) \, \phi_p^2 \, \xi^3 + (6-19 \bar{\phi}) \, \phi_p^2 \, \xi^2 + (2 \phi_p^2 \bar{\phi}+3 \bar{\phi}-6) \, \xi -2 \bar{\phi},
\end{eqnarray}
\begin{eqnarray}
f_2(a) = [\xi \phi_p^2 (1-6\xi)-1][3 \phi_p^2 (\bar{\phi}-2) \xi^2 - (2 \phi_p^2 \bar{\phi}+3 \bar{\phi}-6) \xi + 2 \bar{\phi}],
\end{eqnarray}
\begin{eqnarray}
f_3(a) = [\xi \phi_p^2 (1-6\xi)-1][6 \phi_p^2 \bar{\phi} \xi^2 - 2 ( \phi_p^2 \bar{\phi}+3 \bar{\phi}) \xi + 2 \bar{\phi}],
\end{eqnarray}
\begin{eqnarray}
f_4(a) = \frac{1}{2} [(21 \bar{\phi} - 48) \phi_p^2 \xi^3 + (12-22\bar{\phi}) \phi_p^2 \xi^2 + (4 \phi_p^2 \bar{\phi} + 6 \bar{\phi}-12) \xi - 4 \bar{\phi} ],
\end{eqnarray}
\begin{eqnarray}
f_5(a) = \frac{1}{2} [54 \phi_p^2 \bar{\phi} \xi^3 - 30 \phi_p^2 \bar{\phi} \xi^2 + 4 \bar{\phi} (\phi_p^2+3) \xi - 4 \bar{\phi} ]. 
\end{eqnarray}
\end{widetext}

These expressions can be simplified in the limit $\xi\ll 1$. In this case we obtain,
\begin{equation}\label{s2.13}
\mu \simeq 1 + \phi_p^2 \, \xi,
\end{equation}
\begin{equation}\label{s2.14}
\gamma \simeq 1 - \frac{1}{2} \frac{8 \, k^2 + 9 \, a^2 m^2}{k^2 + a^2 m^2} \phi_p^2 \xi^2.
\end{equation}
As we can see from these results, for a simple quintessence-model, we need a non-minimal coupling in order to have a difference with respect to General Relativity in $\mu$ and $\gamma$ in the sub-Hubble limit. In next section, we will extend this analysis to general  modifications of General Relativity which involve vector degrees of freedom.

\section{General Vector field case}\label{sec3}

In this section we will extend the formalism introduced in previous sections to modified gravities 
involving an extra vector field $A_\mu$. Very much as in 
the scalar case, we will decompose $A_\mu$ in 
a homogeneous background and a perturbation as $A_\mu=A_\mu^{(0)}(\tau)+\delta A_\mu$.
Notice that unlike previous works \cite{DeFelice:2016uil,Lagos}, we will 
allow for the background vector $A_\mu^{(0)}$
to have both non-vanishing temporal and spatial components. For simplicity we will limit ourselves to the case of linearly polarized $A_i^{(0)}(\tau)$.
In such a case,  the background metric
is no longer of the Robertson-Walker type but an axisymmetric Bianchi I metric. This 
metric is characterized by the spatial metric tensor $\Xi_{i j}$
that in general can be written as \cite{Pereira}
\begin{eqnarray}
\Xi_{i j}= e^{2\beta_i(\tau)}\delta_{ij}
\end{eqnarray}
with 
\begin{eqnarray}
\sum_{i=1}^3\beta_i=0 \label{norm}
\end{eqnarray}
This guarantees that $\Xi^{ik}\Xi_{kj}= \delta^i_j$.
Using this metric we can now define a unit spatial vector field 
$\hat A_i$ in the direction of
the background vector field as
\begin{eqnarray}
\hat A_i= \frac{A_i^{(0)}}{A} 
\end{eqnarray}
with $A=(\Xi^{ij}A_i^{(0)} A_j^{(0)})^{1/2}$.

In terms of the unit vector, the spatial metric can be written as
\begin{equation}\label{s3.2}
\Xi_{i j} = C_1 \, \delta_{i j} + C_2 \, \hat{A}_{i} \hat{A}_{j},
\end{equation}
being $C_1$ and $C_2$ functions of time only and satisfying
$C_2=1-C_1^3$ by virtue of (\ref{norm}). This tensor reduces to $\Xi_{i j} = \delta_{i j}$ in the isotropic limit.

The perturbed Bianchi metric in the longitudinal 
gauge, including scalar $(\Phi,\Psi)$, vector $Q_i$ and tensor $h_{ij}$ perturbations reads \cite{Pereira},
\begin{align}\label{s3.1}
ds^{2}=a^{2} \, &\left[ -(1+2\Psi) \, d \tau^{2}+[(1-2\Phi) \, \Xi_{i j} + h_{i j}] \, dx^{i} dx^{j} \right. \nonumber\\
& \left. - 2 \, Q_{i} \, d \tau \, dx^{i} \right],
\end{align}
 Additionally,  the perturbed vector field can be 
 decomposed as $\delta A_{0}$, $\delta A^{\parallel}_{i}$, $\delta A^{\perp}_{i}$, where the different perturbations satisfy in Fourier space
\begin{equation}\label{s3.3}
\hat{k}^{i} \, Q_{i}=0,
\end{equation}
\begin{equation}\label{s3.4}
\hat{k}^{i} \, h_{i j}=0,
\end{equation}
\begin{equation}\label{s3.5}
h_{\,\,\,i}^{i}=0,
\end{equation}
\begin{equation}\label{s3.6}
\hat{k}^{i} \, \delta A^{\perp}_{i}=0,
\end{equation}
\begin{equation}\label{s3.7}
\delta A^{\parallel}_{i}=\delta A^{\parallel} \, \hat{k}_{i},
\end{equation}
being $\hat{k}_{i}$ the unitary direction of the perturbation 
wavevector $k_{i}$ with respect to the spatial metric $\Xi_{ij}$. Notice that indices in spatial vectors
are raised and lowered with the metric $\Xi_{ij}$

As in the scalar field case, we have the perturbed equations,
\begin{equation}\label{s3.8}
\delta \bar{G}^{\mu}_{\,\,\,\nu}=8 \pi G \, \delta T^{\mu}_{\,\,\,\nu},
\end{equation}
\begin{equation}\label{s3.9}
\delta L_{\mu}=0,
\end{equation}
being $\bar{G}^{\mu}_{\,\,\,\nu}$ the modified Einstein tensor and $\delta L_{\mu}$ the perturbed vector field equations. The only matter-energy content we consider is presureless matter as in the scalar field case so that
\begin{eqnarray}
 T^{\mu}_{\,\,\,\nu}=\rho\, u^{\mu}u_{\nu}
\end{eqnarray}
where 
\begin{eqnarray}
\rho=\rho_0+\delta\rho
\end{eqnarray}
and the four-velocity of matter $u^\mu=dx^{\mu}/ds$ is
\begin{eqnarray}
u^\mu=a^{-1}(1-\Psi,v^i)
\end{eqnarray}
so that 
\begin{eqnarray}
u_\mu=a(-1-\Psi,v_i)
\end{eqnarray}
where as mentioned before $v_i=\Xi_{ij}v^j$. The velocity perturbation can
also be decomposed in a longitudinal and transverse (vorticity) components as
\begin{eqnarray}
v_i=v_i^\parallel + v_i^\perp
\end{eqnarray}
such that in Fourier space
\begin{equation}
\hat{k}^{i} \, v^{\perp}_{i}=0,
\end{equation}
\begin{equation}
v^{\parallel}_{i}= v^{\parallel} \, \hat{k}_{i},
\end{equation}

 Additionally, Bianchi 
identities imply the conditions (\ref{s1.6}). Taking such conditions into account, and contracting the spatial components with $\hat k^i$ or $\hat A^i$ we obtain the following set of independent scalar equations: $\delta \bar{G}^{0}_{\,\,0}$, $\hat{A}^{i}\delta \bar{G}^{0}_{\,\,i}$, $\hat{A}^{i}\hat{A}_{j}\delta \bar{G}^{j}_{\,\,i}$ and $\delta \bar{G}^{i}_{\,\,i}$. From the vector field equations we have: $\delta L_{0}$, $\hat{k}^{i} \delta L_{i}$ and $\hat{A}^{i} \delta L_{i}$. Thus we obtain the following seven independent equations,
\begin{align}\label{s3.10}
& \delta \bar{G}^{0}_{\,\,\,0} = -8 \pi G \, \rho \, \delta, \\
& \hat{A}^{i} \delta \bar G^{0}_{\,\,\,i} = -8 \pi G \, \rho \, \hat{A}^{i} v_i, \label{0i}\\
& \hat{A}_{j} \hat{A}^{i} \delta \bar G^{j}_{\,\,\,i} = 0, \label{ij}\\
& \delta \bar G^{i}_{\,\,\,i} = 0, \label{ii}\\
& \delta L_{0} = 0, \label{L0} \\
& \hat{k}^{i} \delta L_{i} = 0, \label{s3.10k} \\
& \hat{A}^{i} \delta L_{i} = 0, \label{s3.10f}
\end{align}
for seven variables: $\Phi$, $\Psi$, $\delta A_0$, $\delta A^{\parallel}$, $\bar{Q} \equiv \hat{A}^{i} Q_{i}$, $\delta \bar{A}^{\perp} \equiv \hat{A}^{i} \delta A_{i}^{\perp}$ and $\bar{h} \equiv \hat{A}^{i} \hat{A}^{j} h_{i j}$. If we apply the QSA, the system (\ref{s3.10})-(\ref{s3.10f}) transforms into an algebraic system for the above variables in terms of the matter variables $\delta$ and $\hat A^i v_i^\perp$ 
(notice that very much as in the scalar case, the scalar velocity perturbation $v^\parallel$ does not contribute to the equations in the QSA),

\begin{widetext}
\begin{eqnarray}\label{s3.11}
A_{1 1} \, \Psi + A_{1 2} \, \Phi + A_{1 3} \, \bar{Q} + A_{1 4} \, \bar{h} 
+A_{1 5}  \, \delta  A_0 
+A_{1 6} \, \delta \hat A^\perp +A_{1 7}  \,\delta  A^\parallel= -4\pi G\,a^2 \rho \delta,
\end{eqnarray}
\begin{eqnarray}\label{s3.12}
A_{2 1} \, \Psi + A_{2 2} \, \Phi + A_{2 3} \, \bar{Q} + A_{2 4} \, \bar{h} 
+ A_{2 5}  \, \delta  A_0
+A_{2 6}  \,\delta \hat A^\perp +A_{2 7} \,\delta  A^\parallel= 16\pi G\,a^2 \rho \hat{A}^{i}v^{\perp}_i,
\end{eqnarray}
%
%
%
%
\begin{eqnarray}\label{s3.14}
A_{m 1} \, \Psi + A_{m 2} \, \Phi + A_{m 3} \, \bar{Q} + A_{m 4} \, \bar{h} 
+ A_{m 5}  \,\delta  A_0
+A_{m 6} \, \delta \hat A^\perp +A_{m 7}  \,\delta  A^\parallel= 0,\;\;\;\; m=3,4,5,6,7
\end{eqnarray}
%

\end{widetext}
where we have introduced a $-4\pi G\,a^2$ factor in (\ref{s3.11}) and $16\pi G\,a^2$ factor in (\ref{s3.12}) for convenience. Here $A_{m n}$ with $m,n=1\dots 7$ are assumed to be arbitrary independent functions of background quantities and $k$ where $k^2=k_i k_j \Xi^{ij}$.
Had we made additional assumptions, such as diffeomorphisms invariance of the starting action,
a simplification of the system of equations would be possible. However the approach we will 
follow in this work is to keep the most general expression for the equations.  
Solving for $\delta A_0$, $\delta A^{\parallel}$ and $\delta \bar{A}^{\perp}$
from (\ref{s3.14}) with $m=5,6,7$ and substituting in the rest of equations
we can obtain each perturbation as a general linear function of $\delta$ and $\hat{A}^{i}v^{\perp}_i$. Restoring indices, we have the following effective equations for the metric perturbations,
\begin{align}\label{s3.15}
k^2 \, \Phi = - 4 \pi G \, a^2 \, \rho \, (\mu_\Phi \, \delta + \eta_\Phi \, \hat{A}^{i} v^{\perp}_i), 
\end{align}
\begin{align}\label{s3.16}
k^2 \, \Psi = - 4 \pi G \, a^2 \, \rho \, (\mu_\Psi \, \delta + \eta_\Psi \, \hat{A}^{i} v^{\perp}_i), 
\end{align}
\begin{align}\label{s3.17}
k^2 \, Q_i = 16 \pi G \, a^2 \, \rho \, (\mu_Q \, \tilde{A}_i \, \delta + \eta_Q \, v^{\perp}_i), 
\end{align}
\begin{align}\label{s3.18}
k^2 \, h_{i j} = - 4 \pi G \, a^2 \, \rho \, (\mu_h \, \Sigma_{i j} \, \delta + \eta_h \, \Lambda_{i j}), 
\end{align}
where, defining $x \equiv \hat{A}^i \hat{k}_i$, we have 
\begin{align}\label{s3.19}
\tilde{A}_i = \hat{A}_j- x \, \hat{k}_j, 
\end{align}
\begin{align}\label{s3.20}
\Sigma_{i j} = 2 \, \tilde{A}_i \, \tilde{A}_j-(1-x^2)(\delta_{i j}-\hat{k}_i \, \hat{k}_j), 
\end{align}
\begin{align}\label{s3.21}
\Lambda_{i j} = 2 \, v^{\perp}_{(i} \, \tilde{A}_{j)} - \frac{(\delta_{i j}-\hat{k}_i \, \hat{k}_j) \, \hat{A}^{k} v^{\perp}_k}{1-x^2}. 
\end{align}
These quantities satisfy the following properties,
\begin{align}\label{s3.22}
\hat{k}^i \tilde{A}_i = 0,
\end{align}
\begin{align}\label{s3.23}
&\Sigma_{i j} = \Sigma_{j i}, \,\,\, \hat{k}^{i} \Sigma_{i j} = 0, \,\,\, \Sigma^i_{\,\,i} = 0,
\end{align}
\begin{align}\label{s3.24}
&\Lambda_{i j} = \Lambda_{j i}, \,\,\, \hat{k}^{i} \Lambda_{i j} = 0, \nonumber\\
&\Lambda^i_{\,\,i} = 0, \,\,\, \hat{A}^{i} \hat{A}^{j} \Lambda_{i j} = \hat{A}^{k} v^{\perp}_k.
\end{align}
With these definitions we see that eight dimensionless parameters 
$(\mu_\Phi, \eta_\Phi,\mu_\Psi,\eta_\Psi, \mu_Q, \eta_Q, \mu_h, \eta_h)$ are needed  to parametrize the theory. In General Relativity they take the values  $(1, 0, 1, 0, 0, 1, 0, 0)$. Notice that if we  consider only a temporal component for the 
background vector field, so that $A_i^{(0)}=0$ then the number of  parameters 
is reduced to three 
($\mu_\Phi, \mu_\Psi, \eta_Q)$, 
in agreement with previous works \cite{DeFelice:2016uil, Lagos}.  In the general case, if dark matter vorticity can be neglected, the number of parameters can be 
reduced from eight to four and, in this case, we can define the
standard  ($\mu,\gamma$) parameters as $\mu=\mu_\Psi$ and $\gamma= \mu_\Phi/\mu_\Psi$. 

In the following we are going to obtain explicit expressions of the complete set of parameters as a function of $k$ and $x$. With that purpose, we will 
firstly derive the dependence in $k$ and $x$ of each $A_{ij}$ coefficient of equations (\ref{s3.11}), (\ref{s3.12}) and (\ref{s3.14}). Notice that on general grounds,  Einstein and field equations can be classified in three categories: scalar equations $\delta S=\{\delta \bar G^{0}_{\,\,\,0}, \delta L_0\}$, vector equations  $\delta V_i=\{\delta \bar G^{0}_{\,\,\,i}, \delta L_i\}$ and tensor equations  $\delta \bar G^{i}_{\,\,\,j}$. These equations have the following general structure taking into account the different linear perturbations. For the left hand side of scalar equations we have
\begin{align}\label{s3.25}
\delta S= &E^{\Phi} \, \Phi+E^{\Psi} \, \Psi+E^{0} \, \delta A_{0} \nonumber\\
 &+E^{Q \, i} \, Q_{i}+E^{\perp \, i} \, \delta A^{\perp}_{i}+E^{\parallel \, i} \, \delta A^{\parallel}_{i} \nonumber\\
 &+E^{i j} \, h_{i j} \, .
\end{align}
For the vector ones
\begin{align}\label{s3.26}
\delta V_{\,\,i}=&E^{\Phi}_{\,\,\, i} \, \Phi+E^{\Psi}_{\,\,\, i} \, \Psi+E^{0}_{\,\,\, i} \, \delta A_{0}  \nonumber\\
 &+E^{Q} \, Q_{i}+E^{\perp} \, \delta A^{\perp}_{i}+E^{\parallel} \, \delta A^{\parallel}_{i}  \nonumber\\
 &+E^{Q \, j}_{\,\,\, i} \, Q_{j}+E^{\perp \, j}_{\,\,\, i} \, \delta A^{\perp}_{j}+E^{\parallel \, j}_{\,\,\, i} \, \delta A^{\parallel}_{\,\,\, j} \nonumber\\
 &+E^{j k}_{\,\,\,\,\,\, i} \, h_{j k} + E^{j} \, h_{j i} \, ,
\end{align}
and for the tensor one
\begin{align}\label{s3.27}
\delta \bar G^{i}_{\,\,\,j}=&E^{\Phi \, i}_{\,\,\, j} \, \Phi+E^{\Psi \, i}_{\,\,\, j} \, \Psi+E^{0 \, i}_{\,\,\, j} \, \delta A_{0} \nonumber\\
&+E^{Q \, i} \, Q_{j}+E^{\perp \, i} \, \delta A^{\perp}_{j}+E^{\parallel \, i} \, \delta A^{\parallel}_{j} \nonumber\\
&+E^{Q}_{\, j} \, Q^{i}+E^{\perp}_{\, j} \, \delta A^{\perp \, i}+E^{\parallel}_{\, j} \, \delta A^{\parallel \, i} \nonumber\\
&+E^{Q \, ik}_{\,\,\, j} \, Q_{k}+E^{\perp \, ik}_{\,\,\, j} \, \delta A^{\perp}_{k}+E^{\parallel \, ik}_{\,\,\, j} \, \delta A^{\parallel}_{k} \nonumber\\
&+E^{T \, i l m}_{j} \, h_{l m} + E^{T \, i l} \, h_{l j} + E^{T \, l}_{\,\,\, j} \, h_{l}^{\,\, i} + E^{T} \, h_{\,\,\,j}^{i}  \, .
\end{align}
Where taking into account the QSA, the $E$ operators are second order differential operators involving only spatial derivatives. Notice also 
that for the scalar and vector equations we have different $E$ operators 
for each equation, but all of them will have the same structure. Thus in Fourier 
space the most general form of the operators are:
\begin{equation}\label{s3.28}
E=A_{1}+A_{2}^{i} \, k_{i}+A_{3} \, k^{2}+A_{4 \, lm} \, k^{l} k^{m},
\end{equation}
\begin{align}\label{s3.29}
E_{i}= & B_{1 \, i}+B_{2 \, i}^{\,\,\, j}\, k_{j}+B_{3} \, k_{i} \nonumber\\
&+B_{4 \, i} \, k^{2} + B_{5 \, l} \, k^{l} k_{i} + B_{6 \, i}^{\,\,\, j k} \, k_{j} k_{k},
\end{align}
\begin{align}\label{s3.30}
E_{j}^{i}=&C_{1 \, j}^{\,\,\, i}+C_{2 \, j} \, k^{i}+C_{3}^{\,\,\, i} \, k_{j}\nonumber\\
&+C_{4 \, jl}^{\,\,\, i} \, k^{l}+C_{5 \, j}^{\,\,\, i} \, k^{2}+C_{6 \, jlm}^{\,\,\, i} \, k^{l} k^{m} \nonumber\\
&+C_{7} \, k^{i} k_{j} + C_{8 \, l}^{i} \, k^{l} k_{j} + C_{9 \, j}^{l} \, k^{i} k_{l},
\end{align}
\begin{align}\label{s3.31}
E_{j}^{ik}=&D_{1 \, j}^{\,\,\, ik}+D_{2 \, j}^{\,\,\, k} \, k^{i}+D_{3 \, j}^{i} \, k^{k}+D_{4}^{\,\,\, ik} \, k_{j}\nonumber\\
&+D_{5 \, jl}^{\,\,\, ik} \, k^{l}+D_{6 \, j}^{\,\,\, ik} \, k^{2}+D_{7 \, jlm}^{\,\,\, ik} \, k^{l} k^{m} \nonumber\\
&+D_{8 \, j} \, k^{i} k^{k}+D_{9}^{\,\,\, i} \, k_{j} k^{k}+D_{10}^{\,\,\,\,\,\, k} \, k^{i} k_{j} \nonumber\\
&+D_{11 \, j l}^{k} \, k^{l} k^{i} + D_{12 \, j l}^{i} \, k^{l} k^{k} + D_{13 \, l}^{i k} \, k^{l} k_{j}.
\end{align}
\begin{align}\label{s3.32}
E^{i l m}_{\,\,\,\, j} =& F^{i l m}_{1 \,\,\,\, j} + F^{i l m n}_{2 \,\,\,\, j} \, k_{n} + F^{i l}_{3 \,\,\,\, j} \, k^{m} \nonumber\\
&+ F^{i m}_{4 \,\,\,\, j} \, k^{l} + F^{l m}_{5 \,\,\,\, j} \, k^{i} + F^{i l m}_{6} \, k_{j} \nonumber\\
&+ F^{i l m k n}_{7 \,\,\,\, j} \, k_{k} k_{n} + F^{i l n}_{8 \,\,\,\, j} \, k_{n} k^{m} + F^{i m n}_{9 \,\,\,\, j} \, k_{n} k^{l} \nonumber\\
&+ F^{l m n}_{10 \,\,\,\, j} \, k_{n} k^{i} + F^{i l m n}_{11} \, k_{n} k_{j} + F^{m}_{12 \,\,\,\, j} \, k^{i} k^{l} \nonumber\\
&+ F^{l}_{13 \,\,\,\, j} \, k^{i} k^{m} + F^{l m}_{14} \, k^{i} k_{j} + F^{i}_{15 \,\,\,\, j} \, k^{l} k^{m} \nonumber\\
&+ F^{i m}_{16} \, k^{l} k_{j} + F^{i l}_{17} \, k^{m} k_{j} + F^{i l m}_{18 \,\,\,\, j} \, k^2 \, ,
\end{align}
where $A$, $B$, $C$, $D$ and $F$ coefficients are in general functions of  background quantities (depending only on time) and their indices  only come from the vector field $A_{i}^{(0)}$ and the $\delta^{i}_{\,\,\, j}$ tensor in all possible combinations.

Once we have the form of the l.h.s of (\ref{s3.10})-(\ref{s3.10f}), we can obtain the most general form of the $A_{ij}$ coefficients  from the scalar equations and from the equations obtained by contracting the vector type equations as $\hat{A}^{i}\delta V_{\,\,i}$ and the tensor one as $\hat{A}^{i}\hat{A}_{j}\delta \bar{G}^{j}_{\,\,i}$. We summarize in Table \ref{ta1}
the general structure of each coefficient.

\begin{table*}[htbp]
\begin{center}
  \begin{tabular}{|c||c|} \hline
$A_{i j}$ for $i$, $j$ = 1, ... 6 & $b_1 + b_2 \, x \, k + (b_3 + b_4 \, x^2) \, k^2$ \\ \hline
$A_{7 i}, A_{i 7}$ for $i$ = 1, ... 6      & $b_1 \, x + (b_2 + b_3 \, x^2) \, k + (b_4 + b_5 \, x^2) \, x \, k^2$ \\ \hline
$A_{7 7}$                      & $b_1 + b_2 \, x^2 + (b_3 + b_4 \, x^2) \, x \, k + (b_5 + b_6 \, x^2 + b_7 \, x^4) \, k^2$ \\ \hline
 \end{tabular}
\caption{Generic structure of the $A_{i j}$ coefficients of the system of equations (\ref{s3.10}). Notice that for every $A_{ij}$ coefficient the corresponding  $b_{\alpha}(a)$ are different functions of time only.}
\label{ta1}
\end{center}
\end{table*}

%
We can now solve the system
of equations (\ref{s3.11}), (\ref{s3.12}) and (\ref{s3.14})  and obtain the coefficients for $\{\Phi, \Psi, \bar{Q}, \bar{h}\}$ in terms of $\delta$ and $\hat{A}^i v^{\perp}_{i}$ which leads to equations (\ref{s3.15})-(\ref{s3.18}).

We have to notice that apart from $k$, the other 
dimensional (comoving) scales appearing in the $E$ operators are the  Hubble rate ${\cal H}$ and 
the mass scale of the vector field. Unlike the usual assumptions in modified
gravities with  the scalar degrees of freedom \cite{Pogosian,Tsujikawa:2008uc}, there is no general argument with which we can determine the dependence of each parameter $A\dots F$ on ${\cal H}$ or on the vector mass. Moreover, the comoving mass scale could be of order ${\cal H}$ due to the background equations. For these reasons,
we cannot a priori neglect any of the $A_{ij}$ terms and we will consider two generic cases: a) we consider the general case in which we keep all the terms in the $E$ operators and b) we assume that all the dimensional
parameters of the $E$ coefficients are of order ${\cal H}$ or ${\cal H}^2$ so that the corresponding
terms can be neglected compared to the $k^2$ terms in the sub-Hubble limit. 
\\
\subsection*{a) General case:}

We find that the following form for the parameters
\begin{eqnarray}\label{s3.33.0}
M(a,k,x) &=& \frac{ \left[P^{14}_{M} (a,k,x) + P^{12}_{M} (a,k,x) \, x k\right] \, k^2}{P^{16}_{D} (a,k,x) + P^{14}_{D} (a,k,x) \, x k} \;\;
\\\mbox{with}\; M&=& \mu_\Phi, \mu_\Psi, \eta_\Phi, \eta_\Psi, \eta_Q, \eta_h \nonumber 
\end{eqnarray}
whereas for $\mu_Q$ and $\mu_h$ we obtain,
\begin{equation}\label{s3.33.1}
\mu_Q(a,k,x) = \frac{ \left[P^{14}_{\mu_Q} (a,k,x) + P^{12}_{\mu_Q} (a,k,x) \, x k\right] \, k^2}{\left[P^{16}_{D} (a,k,x) + P^{14}_{D} (a,k,x) \, x k\right] \, (1-x^2)},
\end{equation}
\begin{equation}\label{s3.33.2}
\mu_h(a,k,x)= \frac{ \left[P^{14}_{\mu_h} (a,k,x) + P^{12}_{\mu_h} (a,k,x) \, x k\right] \, k^2}{\left[P^{16}_{D} (a,k,x) + P^{14}_{D} (a,k,x) \, x k\right] \, (1-x^2)^2},
\end{equation}
where we have defined the following function,
\begin{equation}\label{s3.34}
P^n_{A} (a,k,x) = \sum_{i=1}^{n/2} \, \sum_{j=0}^{i} \, p^{\, (A)}_{i j}(a) \, x^{2 j} \, k^{2i},
\end{equation}
being $p_{i j}^{\, (A)}(a)$ functions of background quantities
and we define $P^n_{A} (a,x)\equiv P^n_{A} (a,1,x) $ which are polynomials of order $n$ with only even powers of $x$. Notice also that
the polynomials in the numerators are in general different for the 
different parameters, whereas those in the  denominators $P^{14}_{D} (a,k,x)$ and $P^{16}_{D} (a,k,x)$ are the same for all of them as they come from the 
inverse of the determinant corresponding to the system of linear equations
\cite{Pogosian}.

Notice that  if we take $x=0$ i.e. we neglect the
anisotropic contributions coming from the spatial components of the
background vector field in equation (\ref{s3.33.0}) we get

\begin{equation}\label{s3.35}
M(a,x,k) = \frac{k^2 \, P_M^{12} (a,k)}{P_D^{14} (a,k)},
\end{equation}
i.e. the ratio of two degree-fourteen polynomials in $k$. 
We could have anticipated that in this case the result should agree with 
that corresponding to two scalar degrees of freedom (which can be identified with $\delta A_0$ and $\delta A^{\parallel}$) given in  (\ref{s1.30}) and (\ref{s1.31}).
But we see that this is not the case because, unlike the scalar case, 
we did not neglect $k$-independent terms in the $E$ expressions. Only if we neglect $k$-independent terms in all equations, except for those of the vector field perturbations in the vector field equations, we recover the scalar field case:
\begin{equation}\label{s3.36}
M(a,x,k) = \frac{P_M^{4} (a,k)}{P_D^{4} (a,k)}.
\end{equation}

\subsection*{b) Coefficients of order ${\cal H}$:}
If we consider the dimensional coefficients of the $E$ operators to be of order ${\cal H}$ and take the sub-Hubble limit, so that only $k^2$ terms survive, we obtain,
\begin{eqnarray}
\label{s3.37}
M(a,x,k)&=& \frac{P_M^{14} (a,x)}{P_D^{16} (a,x)}, 
\\\mbox{for}\;\;\; M&=& \mu_\Phi, \mu_\Psi, \eta_\Phi, \eta_\Psi, \eta_Q, \eta_h\nonumber 
\end{eqnarray}
and 
\begin{eqnarray}
 \mu_Q &=& \frac{P_{\mu_Q}^{14} (a,x)}{(1-x^2)P_D^{16} (a,x)}, 
\\ \mu_h &=& \frac{P_{\mu_h}^{14} (a,x)}{(1-x^2)^2 P_D^{16} (a,x)}.
\end{eqnarray}

As we can see, the expressions are scale independent. If we expand them in multipoles we only have even powers of $x$, odd powers are suppressed in the sub-Hubble limit. Thus, for small anisotropy, $A\ll A_0^{(0)}$, we can find an expansion for any of the eight  parameters (that we denote as $\beta$) of the form,
\begin{equation}\label{s3.38}
\beta(a,x) = \beta_0(a) + \beta_{2}(a) \, x^2 + \beta_{4}(a) \, x^4 + O(x^6),
\end{equation}
where  $\beta_0(a)$ provides the isotropic contribution. 
In particular we can find this kind of expansion for the standard parameter $\mu$ and, without vorticity, also for the parameter $\gamma$. 

\section{Non-minimally coupled vector field}\label{sec4}

In this section we will calculate all the modified gravity parameters using the QSA in a simple example of non-minimally coupled vector field.  We will  consider a particular case of  generalized Proca theory \cite{Minamitsuji:2016ydr},
\begin{align}\label{s4.1}
S =& \int \, d^4x \, \sqrt{-g} \, \left[ \frac{M_p^2}{2} \, (R-2\Lambda) - \frac{1}{4} \, F_{\mu \nu} \, F^{\mu \nu} \right. \nonumber\\
&\left. -(m^2 \, g_{\mu \nu} - \xi \, G_{\mu \nu}) \, A^{\mu} \, A^{\nu} \right],
\end{align}
where $g_{\mu \nu}$ is the metric tensor, $R$ and $G_{\mu \nu}$ are the Ricci scalar and the Einstein tensor, $A_{\mu}$ is the vector field, $\xi$ is the dimensionless non-minimal coupling constant, $m$ is the mass of the vector field, $\Lambda$ the cosmological constant, and $F_{\mu \nu} = \partial_{\mu} A_{\nu} - \partial_{\nu} A_{\mu}$. 

The corresponding energy-momentum tensor reads,

\begin{widetext}
\begin{align}\label{s4.2}
T_{\mu \nu}^{(v)} \,=\, & F_{\mu \rho} \, F_{\nu}^{\,\,\, \rho} - \frac{1}{4} \, g_{\mu \nu} \, F^{\rho \sigma} \, F_{\rho \sigma} + 2 \, m^2 \, \left( A_{\mu} \, A_{\nu} - \frac{1}{2} \, g_{\mu \nu} \, A^{\rho} \, A_{\rho} \right) + \xi \, \left( A^{\rho} \, A_{\rho} \, G_{\mu \nu} + A_{\mu} \, A_{\nu} \, R \right) -  \nonumber\\
& - \, \xi \, g_{\mu \nu} \, \left[ \left( \nabla_{\rho}A^{\rho} \right)^2 - 2 \, \nabla_{\rho} A_{\sigma} \nabla^{\rho} A^{\sigma} + \nabla_{\rho} A_{\sigma} \nabla^{\sigma} A^{\rho} - 2 \, A_{\rho} \, \Box A^{\rho} + 2 \, A^{\rho} \, \nabla_{\rho} \nabla_{\sigma} A^{\sigma} \right] \nonumber\\
& - 2 \, \xi \, \left[ \nabla_{\mu} A_{\rho} \nabla_{\nu} A^{\rho} - \nabla_{\rho} A^{\rho} \nabla_{(\mu} A_{\nu )} - \nabla_{\rho} A_{(\mu} \nabla_{\nu )} A^{\rho} + \nabla_{\rho} A_{\mu} \nabla^{\rho} A_{\nu} + A_{\rho} \nabla_{(\mu} \nabla_{\nu )} A^{\rho} \right. \nonumber\\
& \left. - A^{\rho} \nabla_{\rho} \nabla_{( \mu} A_{\nu )} + A_{( \mu} \Box A_{\nu )} - 2 \, A_{(\mu} \nabla_{\nu)} \nabla_{\sigma} A^{\sigma} + A_{( \mu } \nabla_{\rho} \nabla_{ \nu )} A^{\rho} \right].
\end{align}
\end{widetext}
The total energy-momentum tensor is,
\begin{align}\label{s4.3}
T_{\mu \nu}^{(t)} = T_{\mu \nu}^{(m)} + T_{\mu \nu}^{(v)},
\end{align}
where $T_{\mu \nu}^{(m)}$ is the energy-momentum tensor of matter. 
Finally, the vector field equation of motion is,
\begin{align}\label{s4.4}
\nabla_{\mu} F^{\mu \nu} = 2 \, \left( m^2 \, g^{\mu \nu} - \xi \, G^{\mu \nu} \right) \, A_{\mu}.
\end{align}

We will consider for simplicity the metric (\ref{s3.1}) using the approximation that the background is FRW so that $\Xi_{i j} = \delta_{i j}$. 
As shown before, from the perturbed Einstein and field equations
we obtain the system corresponding to Eqs. (\ref{s3.10})-(\ref{s3.10f}).
We find that in this particular case, in the sub-Hubble limit $\delta A_\parallel$ does not contribute to the equations
and that Eq. (\ref{s3.10k}) is no longer independent, so that we are left with 
six equations for six unknowns. Thus from (\ref{L0}) we get
\begin{align}\label{s4.5}
\delta {\cal A}_0 = \mathcal{C} \, \xi \, (4 \, \Phi - r \, \bar{Q}),
\end{align}
and (\ref{s3.10f}) yields
\begin{align}\label{s4.6}
\delta \bar{{\cal A}}^{\perp} = 2 \, \mathcal{C} \, \xi \, r \, (1-x^2) \, (\Phi - \Psi) + \mathcal{C} \, \xi \, (\bar{Q} + r \, \bar{h}),
\end{align}
where
\begin{equation}\label{s4.7}
\mathcal{C} \equiv \frac{A_0^{(0)} (\tau)}{a (\tau) \, M_p}, \,\,\,\,\,\, r \equiv \frac{A(\tau)}{A_0^{(0)}(\tau)}, 
\end{equation}
\begin{equation}\label{s4.7b}
\delta{\cal A}_0 \equiv \frac{\delta A_0 (\tau)}{a (\tau) \, M_p}, \,\,\,\,\,\, \delta \bar{{\cal A}}^{\perp} \equiv \frac{\delta \bar{{A}}^{\perp}}{a (\tau) \, M_p}, 
\end{equation}
and in this case $A(\tau)=(\delta^{ij}A_i^{(0)} A_j^{(0)})^{1/2}$.

The modified Einstein equations provide from (\ref{s3.10})
\begin{align}
&\left[ 1 - \xi \, \mathcal{C}^2 \, (1-r^2 \, x^2) \right] \, k^2 \, \Phi + \xi \, \mathcal{C} \, r \, k^2 \, \delta \bar{{\cal A}}^{\perp} \nonumber\\
&- \frac{1}{2} \, \xi \, \mathcal{C}^2 \, r^2 \, k^2 \, \bar{h} = - 4 \pi G \, a^2 \, \rho \, \delta,
\end{align}
from  (\ref{0i})
\begin{align}\label{s4.8}
&\left[ 1 - \xi \, \mathcal{C}^2 \, (1-r^2) \right] \, k^2 \, \bar{Q} - 4 \, \xi \, \mathcal{C}^2 \, r \, (1-x^2) \, k^2 \, \Phi \nonumber\\
&- 2 \, \xi \, \mathcal{C}^2 \, r \, k^2 \, \bar{h} + 2 \, \xi \, \mathcal{C} \, r \, (1-x^2) \, k^2 \, \delta {\cal A}_0 \nonumber\\
&+ 2 \, \xi \, \mathcal{C} \, k^2 \, \delta \bar{{\cal A}}^{\perp} = 16 \pi G \, a^2 \, \rho \, \hat{A}^{i} v^{\perp}_i,
\end{align}
whereas (\ref{ii}) gives
\begin{align}\label{s4.9}
&\left[ 1 + \xi \, \mathcal{C}^2 \, (1+r^2) \right] \, \Phi - \left[ 1 - \xi \, \mathcal{C}^2 \, (1-r^2 \, x^2) \right] \, \Psi \nonumber\\
&- \xi \, \mathcal{C}^2 \, r \, \bar{Q} + \xi \, \mathcal{C} \, r \, \delta \bar{{\cal A}}^{\perp} - 2 \, \xi \, \mathcal{C} \, \delta {\cal A}_0 =0,
\end{align}
and finally (\ref{ij}) yields
\begin{align}\label{s4.10}
&\left[ 1 + \xi \, \mathcal{C}^2 \, (1+r^2) \right] \, \Phi - \left[ 1 - \xi \, \mathcal{C}^2 \, (1-r^2) \right] \, \Psi \nonumber \\
&+ \frac{\left[ 1 + \xi \, \mathcal{C}^2 \, (1+r^2) \right]}{2 \, (1-x^2)} \, \bar{h} - 2 \, \xi \, \mathcal{C} \, \delta {\cal A}_0 = 0. 
\end{align}
Substituting the vector field perturbations from Eqs. (\ref{s4.5}) and (\ref{s4.6})  into the Einstein equations  we can obtain,
\begin{align}\label{s4.11}
k^2 \, \Phi = - 4 \pi G \, a^2 \, \rho \, (\mu_\Phi \, \delta + \eta_\Phi \, \hat{A}^{i} v^{\perp}_i), 
\end{align}
\begin{align}\label{s4.12}
k^2 \, \Psi = - 4 \pi G \, a^2 \, \rho \, (\mu_\Psi \, \delta + \eta_\Psi \, \hat{A}^{i} v^{\perp}_i), 
\end{align}
\begin{align}\label{s4.13}
k^2 \, \bar{Q} = 16 \pi G \, a^2 \, \rho \, (\mu_Q \, (1-x^2) \, \delta + \eta_Q \, \hat{A}^{i} v^{\perp}_i), 
\end{align}
\begin{align}\label{s4.14}
k^2 \, \bar{h} = - 4 \pi G \, a^2 \, \rho \, (\mu_h \, (1-x^2)^2 \, \delta + \eta_h \, \hat{A}^{i} v^{\perp}_i). 
\end{align}
Considering the background equations, ${\cal C}^2 a^2m^2=\Od({\cal H}^2)$, so that 
we are in the case b) analyzed before where the mass terms can be neglected and accordingly  the modified gravity parameters are scale invariant and their general structure  is
\begin{equation}\label{s4.15}
\mu_\Phi(a,x) = \frac{P^4_{\mu_\Phi} (a,x)}{P^4_D (a,x)}, \,\,\,\,\,\,\, \eta_\Phi(a,x) = \frac{P^4_{\eta_\Phi} (a,x)}{P^4_D (a,x)}, 
\end{equation}
\begin{equation}\label{s4.16}
\mu_\Psi(a,x) = \frac{P^2_{\mu_\Psi} (a,x)}{P^4_D (a,x)}, \,\,\,\,\,\,\, \eta_\Psi(a,x) = \frac{P^2_{\eta_\Psi} (a,x)}{P^4_D (a,x)}, 
\end{equation}
\begin{equation}\label{s4.17}
\mu_Q(a,x) = \frac{P^2_{\mu_Q} (a,x)}{P^4_D (a,x)}, \,\,\,\,\,\,\, \eta_Q(a,x) = \frac{P^4_{\eta_Q} (a,x)}{P^4_D (a,x)}, 
\end{equation}
\begin{equation}\label{s4.18}
\mu_h(a,x) = \frac{P^2_{\mu_h} (a,x)}{P^4_D (a,x)}, \,\,\,\,\,\,\, \eta_h(a,x) = \frac{P^6_{\eta_h} (a,x)}{P^4_D (a,x)}. 
\end{equation}

The reduced order of the polynomials  with respect to the general 
result in (\ref{s3.33.0}), (\ref{s3.33.1}) and (\ref{s3.33.2})
 is due to the fact that in this particular case we do not have all the possible $x$ dependence in each coefficient.
 
The explicit expression of the different polynomials is not very informative 
and we do not include it here. However in the limit $r\ll 1$ (small anisotropy) and $\xi\ll 1$ we can obtain simple expressions 
 to next to leading order in both parameters 
\begin{equation}\label{s4.19}
\mu_\Phi(a,x) = 1 + \xi \, \mathcal{C}^2 \, (1-r^2 \, x^2), 
\end{equation}
\begin{equation}\label{s4.20}
\eta_\Phi(a,x) = 4 \, \xi^2 \, \mathcal{C}^2 \, r , 
\end{equation}
\begin{equation}\label{s4.21}
\mu_\Psi(a,x) = 1 + 3 \, \xi \, \mathcal{C}^2 + \xi \, \mathcal{C}^2 \, r^2 \, (1-2\,x^2) , 
\end{equation}
\begin{equation}\label{s4.22}
\eta_\Psi(a,x) = 4 \, \xi \, \mathcal{C}^2 \, r , 
\end{equation}
\begin{equation}\label{s4.23}
\mu_Q(a,x) = -\,\xi \, \mathcal{C}^2 \, r , 
\end{equation}
\begin{equation}\label{s4.24}
\eta_Q(a,x) = 1+\xi \, \mathcal{C}^2 (1-r^2), 
\end{equation}
\begin{equation}\label{s4.25}
\mu_h(a,x) = 2 \, \xi \, \mathcal{C}^2 \, r^2 , 
\end{equation}
\begin{equation}\label{s4.26}
\eta_h(a,x) = 8 \, \xi \, \mathcal{C}^2 \, r \, (1-x^2). 
\end{equation}
As expected we recover the General Relativity values when $\xi=0$ as in the scalar field case and the sub-Hubble limit.  In the absence of vorticity, we can define $\gamma = \mu_\Phi / \mu_\Psi$ as mentioned before so
that,
\begin{equation}\label{s4.27}
\gamma(a,x) = 1 - 2 \, \xi \, \mathcal{C}^2 - \xi \, \mathcal{C}^2 \, r^2 \, (1-x^2) .
\end{equation}
%

\section{Anisotropic effects on galaxy clustering and weak lensing}\label{sec5}

In this last section we will study the potential observable effects of the  anisotropic modified gravity parameters in galaxy surveys. For simplicity, we will consider only scalar perturbations in the sub-Hubble regime (case b) above ) and negligible dark matter vorticity, so that we are left with only 
two effective parameters which are scale invariant,  $\mu = \mu (a,x)$ and $\gamma = \gamma (a,x)$. We will also assume that  the background 
vector field is a subdominant contribution with respect to 
matter so that the background evolution can be correctly described by 
a Robertson-Walker metric, i.e. $\Xi_{ij}=\delta_{ij}$. In this 
framework, the growth equation for pressureless matter is \cite{DE},
\begin{eqnarray}\label{s5.1}
\ddot{\delta}_m+\left(2+\frac{\dot{H}}{H}\right)\dot{\delta}_m-\frac{3}{2}\,\mu(a,x)\,\Omega_m(a)\,\delta_m\simeq 0 \label{growth},
\end{eqnarray}
where dots denotes derivative with respect to $\ln a$, ${\cal H}=aH$ and $\Omega_{m} (a)$ is the matter density parameter $\Omega_{m} (a)=\Omega_m \, a^{-3} \, \frac{H_{0}^{2}}{H^{2}(a)}$.
   By solving this equation we obtain the growth factor $D(z,x) = \delta_m (z,x) / \delta_m (0)$, where as usual the redshift is related to the scale 
   factor by $a=1/(1+z)$, and the growth function $f(z,x) = \dot{D}(z,x)/D(z,x)$ which unlike in the ordinary General Relativity 
case is anisotropic because of the $x$ dependence.  Then, the redshift-space galaxy power spectrum is \cite{DE},
\begin{equation}\label{s5.2}
P_g(k,\hat{\mu},x,z)= \left(1+ \, \frac{f(z,x)}{b} \, \hat{\mu}^{2}\right)^{2} \,\, D^2(z,x) \,\, b^2 \,\, P_m(k),
\end{equation}
where $b$ is the galaxy bias, $P_m(k)$ is the matter power spectrum and $\hat{\mu} = \hat{n}_i \hat{k}^i$ with $\hat{n}_i$  pointing along the line of sight. As we can see, the redshift space galaxy power spectrum has two anisotropic contributions: on one hand the standard contribution from redshift space distorsions (RSD) which introduces a quadrupole and hexadecapole
in $\hat \mu$ with $m=0$ in the spherical harmonic expansion, and on the other, 
an extra contribution coming from the $x$ dependence of the growth function. This new contribution introduces arbitrary-order multipole contributions with 
$m \neq 0$. This fact could help to discriminate the modified gravity contribution from the standard RSD effect.

We can also consider the weak lensing  power spectra. In this situation, we have to take into account that we transform to the Fourier space only the space transverse  to the line of sight, whose coordinates are denoted as $\vec{\theta} = (\theta_1,\, \theta_2)$. The corresponding conjugate coordinates in Fourier space are $\vec{\ell}=(\ell_1,\, \ell_2)$. We now choose the $z$ axis along the 
line of sight, so that $\hat A^i$ with $i=1,2$ is the projection of the 
unit vector on the transverse space. Thus we can define the cosine of
the angle between the projected vector and $\vec\ell$ as 
\begin{align}\label{s1.72}
\Upsilon \equiv \frac{\hat{A}^i \ell_i}{\ell \, \sqrt{1-\hat{A}_3^2}},
\end{align}
where $\hat{A}_3$ is the projection along the line of sight. In this case  we
have $\ell_1 = \ell \, \Upsilon$ and $\ell_2 = \ell \, \sqrt{1-\Upsilon^2}$. Finally for the power spectra of convergence and shear \cite{DE} we find,
\begin{widetext}
\begin{equation}\label{s5.3}
P_{\kappa}(\ell, \Upsilon) = \frac{9 \, H_0^4 \, \Omega_m^{2}}{4} \, \int^{\infty}_{0} \, \frac{dz}{H(z)} \, \frac{(1+\gamma(z,\Upsilon))^2}{4} \, \mu^2(z,\Upsilon) \, (1+z)^2 \, g^2(z) \, D^2(z,\Upsilon) \, P_m\left(\frac{\ell}{\pi \chi (z)}\right),
\end{equation}
\begin{equation}\label{s5.4}
P_{\gamma_1}(\ell, \Upsilon) = [1-4 \, \Upsilon^2 \, (1-\Upsilon^2)] \, P_{\kappa}(\ell, \Upsilon),
\end{equation}
\begin{equation}\label{s5.5}
P_{\gamma_2}(\ell, \Upsilon) = 4 \, \Upsilon^2 \, (1-\Upsilon^2) \, P_{\kappa}(\ell, \Upsilon),
\end{equation}
\end{widetext}
where $H_0$ is the Hubble parameter today, $g(z)$ is the window function,
\begin{equation}\label{s5.6}
g(z) = \int_z^{\infty} \, \left( 1-\frac{\chi (z)}{\chi (z')} \right) \, n(z') \, dz',
\end{equation}
$n(z)$ is the galaxy density function, and $\chi (z)$ is the comoving radial distance,
\begin{equation}\label{s5.7}
\chi (z) = \int_0^{z} \, \frac{dz'}{H(z')}.
\end{equation}
As we can see, the presence of the vector field introduces an angular dependence in the convergence and shear power spectra so that from weak lensing surveys it would be also  
possible to measure (or constrain) the contributions from this kind of 
modified gravity theories. Additionally,  had we considered vector and tensor perturbations in addition to the scalar ones, the power spectra of convergence and shear would also depend
on $\mu_Q$ and $\mu_h$.  This means that in the general case without vorticity, we could use four independent observables: $P_g$, $P_{\kappa}$, $P_{\gamma_1}$, $P_{\gamma_2}$ to measure four modified gravity parameters: $\mu_\Phi$, $\mu_\Psi$, $\mu_Q$ and $\mu_h$. Finally, in this situation weak lensing  produces and additional rotation effect \cite{Cooray:2002mj}. This effect is also known to be present in  standard $\Lambda$CDM cosmology to second order in perturbation theory. Unfortunately, rotation  cannot be measured  using standard weak lensing surveys because of the lack of information about the original orientation of the galaxy images \cite{Thomas:2016xhb}.

\section{Conclusions}\label{sec6}

In this work we have studied the problem of parametrizing modified gravity theories that involve additional vector degrees of freedom.  Unlike previous
works we have allowed for the presence of both temporal and spatial 
components in the background vector field, thus introducing a  preferred
direction in the equations for perturbations. We have followed a phenomenological (model independent) approach and considered the most general 
modification of gravity with second order equations of motion. We have limited ourselves 
to the case of sub-Hubble modes in the quasi-static approximation.  The main result of this work is that eight effective parameters are needed to describe a general theory in this regime, in contrast with the two parameters
needed in the standard scalar case. In addition, such parameters exhibit and extra
dependence on the direction of the wavevector. In the case in which dark matter
vorticity is negligible, the number of independent parameters is reduced to four. 

We have also obtained the general dependence of the parameters with $x$ and $k$
and showed that  such a dependence is different from that obtained
in the case of a modification of gravity with two scalar degrees of freedom. 
In the case in which the mass scale of the vector field is of the order of the Hubble parameter, we have shown that the effective parameters are scale invariant
to leading order in the sub-Hubble regime and that in this regime the odd multipoles are suppressed in the angular dependence. Thus, in this case a particularly simple expression can be obtained for any of the parameters provided  the spatial anisotropy is small
\begin{equation}
\beta(a,x) = \beta_0(a) + \beta_{2}(a) \, x^2 + \beta_{4}(a) \, x^4 + O(x^6).
\end{equation}
Finally we have explored the possible observational consequences of this
kind of modification of gravity. We have shown that in the case in which we 
only consider the effects of scalar perturbations, the growth function
of matter fluctuations becomes anisotropic and this anisotropy is translated to 
the observables, namely the galaxy power spectrum in redshift space and 
the lensing (convergence and shear) spectra. In the first case, we find that 
unlike the anisotropic contribution from RSD, the vector degrees of freedom 
introduce angular dependences with arbitrary multipolar order and with $m\neq 0$. This fact could help to discriminate both effects in future galaxy survey data.
 
In addition, when the effects of vector and tensor perturbations are taken into account (in the 
vanishing vorticity limit) the four effective parameters $\mu_\Phi$, $\mu_\Psi$,
$\mu_Q$ and $\mu_h$ could be measured from the four power spectra $P_g$, 
$P_\kappa$, $P_{\gamma_1}$ and $P_{\gamma_2}$ thus  providing a
way to test vector modifications of gravity in future surveys \cite{ApMa}.

\vspace{0.2cm}
{\it Acknowledgements}. M.A.R acknowledges support from UCM predoctoral grant. This work has been supported by the Spanish  MINECO grants FIS2016-78859-P(AEI/FEDER, UE) and Red Consolider MultiDark FPA2017-90566-REDC.



\begin{thebibliography}{99}
\bibitem{JPAS} 
  N.~Benitez {\it et al.} [J-PAS Collaboration],
  arXiv:1403.5237 [astro-ph.CO].
\bibitem{DESI} 
  A.~Aghamousa {\it et al.} [DESI Collaboration],
  arXiv:1611.00036 [astro-ph.IM].
  
  \bibitem{Euclid}
  R.~Laureijs {\it et al.} [EUCLID Collaboration],
  arXiv:1110.3193 [astro-ph.CO].

\bibitem{TAIPAN}
  E.~da Cunha {\it et al.},
  Publ.\ Astron.\ Soc.\ Austral.\  {\bf 34} (2017) 47
  doi:10.1017/pasa.2017.41
  [arXiv:1706.01246 [astro-ph.GA]].

\bibitem{LSST} 
  P.~A.~Abell {\it et al.} [LSST Science and LSST Project Collaborations],
  arXiv:0912.0201 [astro-ph.IM].

\bibitem{SKA} 
  C.~Howlett, L.~Staveley-Smith and C.~Blake,
  Mon.\ Not.\ Roy.\ Astron.\ Soc.\  {\bf 464} (2017) no.3,  2517
  doi:10.1093/mnras/stw2466
  [arXiv:1609.08247 [astro-ph.CO]].
\bibitem{forecast} 
  L.~Amendola {\it et al.},
  Living Rev.\ Rel.\  {\bf 21} (2018) no.1,  2
  doi:10.1007/s41114-017-0010-3
  [arXiv:1606.00180 [astro-ph.CO]].

\bibitem{DE}
L. Amendola and S. Tsujikawa, {\it Dark energy: theory and observations}, Cambridge (2010)

\bibitem{Clifton}
  T.~Clifton, P.~G.~Ferreira, A.~Padilla and C.~Skordis,
  Phys.\ Rept.\  {\bf 513} (2012) 1
  doi:10.1016/j.physrep.2012.01.001
  [arXiv:1106.2476 [astro-ph.CO]].
  
  \bibitem{Pogosian}
  A.~Silvestri, L.~Pogosian and R.~V.~Buniy,
  Phys.\ Rev.\ D {\bf 87} (2013) no.10,  104015
  doi:10.1103/PhysRevD.87.104015
  [arXiv:1302.1193 [astro-ph.CO]].
  
  \bibitem{Hordenski}
  J.~Gleyzes, D.~Langlois, F.~Piazza and F.~Vernizzi,
  Phys.\ Rev.\ Lett.\  {\bf 114} (2015) no.21,  211101
  doi:10.1103/PhysRevLett.114.211101
  [arXiv:1404.6495 [hep-th]].
  
  
  \bibitem{HordParam0}
  A.~De Felice, R.~Kase and S.~Tsujikawa,
  Phys.\ Rev.\ D {\bf 83} (2011) 043515
  doi:10.1103/PhysRevD.83.043515
  [arXiv:1011.6132 [astro-ph.CO]].
  
  \bibitem{HordParam1}
  J.~Gleyzes, D.~Langlois and F.~Vernizzi,
  Int.\ J.\ Mod.\ Phys.\ D {\bf 23} (2015) no.13,  1443010
  doi:10.1142/S021827181443010X
  [arXiv:1411.3712 [hep-th]].
  
  \bibitem{HordParam2}
  S.~Tsujikawa,
  Phys.\ Rev.\ D {\bf 92} (2015) no.4,  044029
  doi:10.1103/PhysRevD.92.044029
  [arXiv:1505.02459 [astro-ph.CO]].

\bibitem{Ford}
  L.~H.~Ford,
  Phys.\ Rev.\ D {\bf 40} (1989) 967.
  doi:10.1103/PhysRevD.40.967
  
\bibitem{Golovnev}
  A.~Golovnev, V.~Mukhanov and V.~Vanchurin,
  JCAP {\bf 0806} (2008) 009
  doi:10.1088/1475-7516/2008/06/009
  [arXiv:0802.2068 [astro-ph]].
  
  \bibitem{Koivisto}
  T.~Koivisto and D.~F.~Mota,
  JCAP {\bf 0808} (2008) 021
  doi:10.1088/1475-7516/2008/08/021
  [arXiv:0805.4229 [astro-ph]].
  
  
  \bibitem{Joda1}
  M.~a.~Watanabe, S.~Kanno and J.~Soda,
  Phys.\ Rev.\ Lett.\  {\bf 102}, 191302 (2009)
  doi:10.1103/PhysRevLett.102.191302
  [arXiv:0902.2833 [hep-th]].
  
  \bibitem{Joda2}
  S.~Kanno, J.~Soda and M.~a.~Watanabe,
  JCAP {\bf 1012}, 024 (2010)
  doi:10.1088/1475-7516/2010/12/024
  [arXiv:1010.5307 [hep-th]]

\bibitem{ArmendarizPicon}
  C.~Armendariz-Picon,
  JCAP {\bf 0407} (2004) 007
  doi:10.1088/1475-7516/2004/07/007
  [astro-ph/0405267].
  
\bibitem{Boehmer}
  C.~G.~Boehmer and T.~Harko,
  Eur.\ Phys.\ J.\ C {\bf 50} (2007) 423
  doi:10.1140/epjc/s10052-007-0210-1
  [gr-qc/0701029].  
  
\bibitem{Beltran1}
  J.~Beltran Jimenez and A.~L.~Maroto,
  Phys.\ Rev.\ D {\bf 78} (2008) 063005
  doi:10.1103/PhysRevD.78.063005
  [arXiv:0801.1486 [astro-ph]].  
\bibitem{Beltran2}
  J.~Beltran Jimenez and A.~L.~Maroto,
  JCAP {\bf 0903} (2009) 016
  doi:10.1088/1475-7516/2009/03/016
  [arXiv:0811.0566 [astro-ph]].  
\bibitem{Beltran3}
  J.~Beltran Jimenez and A.~L.~Maroto,
  Phys.\ Lett.\ B {\bf 686} (2010) 175
  doi:10.1016/j.physletb.2010.02.038
  [arXiv:0903.4672 [astro-ph.CO]].  
  
  
  
  
  \bibitem{Redondo}
  P.~Arias, D.~Cadamuro, M.~Goodsell, J.~Jaeckel, J.~Redondo and A.~Ringwald,
  JCAP {\bf 1206} (2012) 013
  doi:10.1088/1475-7516/2012/06/013
  [arXiv:1201.5902 [hep-ph]].
  
  \bibitem{Dimopoulos}
  K.~Dimopoulos,
  Phys.\ Rev.\ D {\bf 74} (2006) 083502
  doi:10.1103/PhysRevD.74.083502
  [hep-ph/0607229].
  
  \bibitem{Nelson}
  A.~E.~Nelson and J.~Scholtz,
  Phys.\ Rev.\ D {\bf 84} (2011) 103501
  doi:10.1103/PhysRevD.84.103501
  [arXiv:1105.2812 [hep-ph]].

\bibitem{ULV}
  J.~A.~R.~Cembranos, A.~L.~Maroto and S.~J.~N\'u\~nez Jare\~no,
  JHEP {\bf 1702} (2017) 064
  doi:10.1007/JHEP02(2017)064
  [arXiv:1611.03793 [astro-ph.CO]].
  
  \bibitem{BGP}
 J.~Beltran Jimenez and L.~Heisenberg,
  Phys.\ Lett.\ B {\bf 757} (2016) 405
  doi:10.1016/j.physletb.2016.04.017
  [arXiv:1602.03410 [hep-th]].
  
  \bibitem{BGP2}
  J.~Beltran Jimenez and L.~Heisenberg,
  Phys.\ Lett.\ B {\bf 770} (2017) 16
  doi:10.1016/j.physletb.2017.03.002
  [arXiv:1610.08960 [hep-th]].

\bibitem{DeFelice:2016uil}
  A.~De Felice, L.~Heisenberg, R.~Kase, S.~Mukohyama, S.~Tsujikawa and Y.~l.~Zhang,
  Phys.\ Rev.\ D {\bf 94} (2016) no.4,  044024
  doi:10.1103/PhysRevD.94.044024
  [arXiv:1605.05066 [gr-qc]].
  
  \bibitem{Lagos}
  M.~Lagos, E.~Bellini, J.~Noller, P.~G.~Ferreira and T.~Baker,
  JCAP {\bf 1803} (2018) no.03,  021
  doi:10.1088/1475-7516/2018/03/021
  [arXiv:1711.09893 [gr-qc]].
  
  \bibitem{isotropyPlanck}
  P.~A.~R.~Ade {\it et al.} [Planck Collaboration],
  Astron.\ Astrophys.\  {\bf 594} (2016) A16
  doi:10.1051/0004-6361/201526681
  [arXiv:1506.07135 [astro-ph.CO]].
  
  \bibitem{anomalies}
  D.~J.~Schwarz, C.~J.~Copi, D.~Huterer and G.~D.~Starkman,
  Class.\ Quant.\ Grav.\  {\bf 33} (2016) no.18,  184001
  doi:10.1088/0264-9381/33/18/184001
  [arXiv:1510.07929 [astro-ph.CO]].
  
  \bibitem{Campanelli}
  L.~Campanelli, P.~Cea and L.~Tedesco,
  Phys.\ Rev.\ Lett.\  {\bf 97} (2006) 131302
   Erratum: [Phys.\ Rev.\ Lett.\  {\bf 97} (2006) 209903]
  doi:10.1103/PhysRevLett.97.131302, 10.1103/PhysRevLett.97.209903
  [astro-ph/0606266].
  
  \bibitem{MDE}
  J.~Beltran Jimenez and A.~L.~Maroto,
  Phys.\ Rev.\ D {\bf 76} (2007) 023003
  doi:10.1103/PhysRevD.76.023003
  [astro-ph/0703483].
  
  \bibitem{ACW}
  L.~Ackerman, S.~M.~Carroll and M.~B.~Wise,
  Phys.\ Rev.\ D {\bf 75} (2007) 083502
   Erratum: [Phys.\ Rev.\ D {\bf 80} (2009) 069901]
  doi:10.1103/PhysRevD.75.083502, 10.1103/PhysRevD.80.069901
  [astro-ph/0701357].
  
  
  \bibitem{Pullen}
  A.~R.~Pullen and C.~M.~Hirata,
  JCAP {\bf 1005} (2010) 027
  doi:10.1088/1475-7516/2010/05/027
  [arXiv:1003.0673 [astro-ph.CO]].

\bibitem{Jeong}
  D.~Jeong and M.~Kamionkowski,
  Phys.\ Rev.\ Lett.\  {\bf 108} (2012) 251301
  doi:10.1103/PhysRevLett.108.251301
  [arXiv:1203.0302 [astro-ph.CO]].
  
  \bibitem{Shiraishi}
  M.~Shiraishi, N.~S.~Sugiyama and T.~Okumura,
  Phys.\ Rev.\ D {\bf 95} (2017) no.6,  063508
  doi:10.1103/PhysRevD.95.063508
  [arXiv:1612.02645 [astro-ph.CO]].
  
  \bibitem{Durrer2}
  V.~Tansella, C.~Bonvin, G.~Cusin, R.~Durrer, M.~Kunz and I.~Sawicki,
  arXiv:1807.00731 [astro-ph.CO].
  
  \bibitem{Bartolo}
  N.~Bartolo, S.~Matarrese, M.~Peloso and A.~Ricciardone,
  Phys.\ Rev.\ D {\bf 87} (2013) no.2,  023504
  doi:10.1103/PhysRevD.87.023504
  [arXiv:1210.3257 [astro-ph.CO]].
  
  \bibitem{Bartolo2}
  N.~Bartolo, S.~Matarrese, M.~Peloso and M.~Shiraishi,
  JCAP {\bf 1507} (2015) no.07,  039
  doi:10.1088/1475-7516/2015/07/039
  [arXiv:1505.02193 [astro-ph.CO]].
  
  \bibitem{Joda3}
  M.~a.~Watanabe, S.~Kanno and J.~Soda,
  Prog.\ Theor.\ Phys.\  {\bf 123}, 1041 (2010)
  doi:10.1143/PTP.123.1041
  [arXiv:1003.0056 [astro-ph.CO]]
  
  \bibitem{Bartolo3}
  N.~Bartolo, A.~Kehagias, M.~Liguori, A.~Riotto, M.~Shiraishi and V.~Tansella,
  Phys.\ Rev.\ D {\bf 97} (2018) no.2,  023503
  doi:10.1103/PhysRevD.97.023503
  [arXiv:1709.05695 [astro-ph.CO]].
  
\bibitem{Tsujikawa:2008uc}
  S.~Tsujikawa, K.~Uddin, S.~Mizuno, R.~Tavakol and J.~Yokoyama,
  Phys.\ Rev.\ D {\bf 77} (2008) 103009
  doi:10.1103/PhysRevD.77.103009
  [arXiv:0803.1106 [astro-ph]].
	
  \bibitem{Pereira}
  T.~S.~Pereira, C.~Pitrou and J.~P.~Uzan,
  JCAP {\bf 0709} (2007) 006
  doi:10.1088/1475-7516/2007/09/006
  [arXiv:0707.0736 [astro-ph]].

\bibitem{Minamitsuji:2016ydr}
  M.~Minamitsuji,
  Phys.\ Rev.\ D {\bf 94} (2016) no.8,  084039
  doi:10.1103/PhysRevD.94.084039
  [arXiv:1607.06278 [gr-qc]].
	
\bibitem{Cooray:2002mj}
  A.~Cooray and W.~Hu,
  Astrophys.\ J.\  {\bf 574} (2002) 19
  doi:10.1086/340892
  [astro-ph/0202411].
	
\bibitem{Thomas:2016xhb}
  D.~B.~Thomas, L.~Whittaker, S.~Camera and M.~L.~Brown,
  Mon.\ Not.\ Roy.\ Astron.\ Soc.\  {\bf 470} (2017) no.3,  3131
  doi:10.1093/mnras/stx1468
  [arXiv:1612.01533 [astro-ph.CO]].
  
\bibitem{ApMa} M. Aparicio Resco, A.L. Maroto, work in progress

%
%
%
%
%
%
%
%
%
%
%
%
%
%
%


\end{thebibliography}
\end{document}